\documentstyle[preprint,aps]{revtex}
\begin{document}
\draft
\preprint{MKPH-T-94-3}
\title{The $\eta$NN Coupling in Eta Photoproduction}
\author{L. Tiator}
\address{Institut f\"ur Kernphysik, Universit\"at Mainz, 55099 Mainz, Germany}
\author{C. Bennhold}
\address{Center of Nuclear Studies, Department of Physics,
The George Washington University, Washington, D.C.
20052, USA}
\author{S. S. Kamalov}
\address{Laboratory of Theoretical Physics, JINR Dubna, Head Post Office
Box 79, 101000 Moscow, Russia.}
\date{\today}
\maketitle
\begin{abstract}
Low-energy eta photoproduction on the nucleon is studied in an effective
Lagrangian approach that contains Born terms, vector meson and nucleon
resonance contributions. The resonance sector includes the $S_{11}(1535)$,
$P_{11}(1440)$ and $D_{13}(1520)$ states whose couplings are fixed by
independent electromagnetic and hadronic data. The available $(\gamma,\eta)$
data are employed to discuss the difference between pseudoscalar and
pseudovector Born terms and to determine the magnitude of the $\eta NN$
coupling constant. We present multipoles that are most sensitive to the various
model ingredients and demonstrate how these multipoles may be accessed in
polarization observables. Cross section calculations are presented for
eta photoproduction on light nuclei.
\end{abstract}
\pacs{PACS numbers: 13.60.Le, 13.60.Rj, 14.40.Aq}

\newpage
\section{INTRODUCTION}
Over the last several years there has been renewed interest in the production
of $\eta$--mesons with protons, pions and electrons and their interaction with
nucleons and nuclei. One of the first $\eta$--nuclear experiments, performed at
SATURNE in 1988\cite{Ber88}, reported surprisingly large eta production rates
 near
threshold in the reaction d(p,$\eta$)$^3$He. These large cross sections
permitted not only a more precise determination of the $\eta$--mass\cite{Plo92}
but were also used to perform rare decay measurements of the eta\cite{Kes93}.
Additional experiments involving pion induced eta production were performed
at Los Alamos\cite{Pen89}. Again, the experimental cross sections at threshold
region of the reaction $^3$He($\pi^{-},\eta$)$^3$H are above the theoretical
calculations\cite{Pen89,Kam93}.

The advent of high duty-cycle electron accelerators opens for the first time
the opportunity to study the reactions $N(\gamma,\eta)N$ and $N(e,e'\eta)N$
in greater detail. Our present knowledge of the $(\gamma,\eta)$ process is
based solely on some old measurements around 20 years ago\cite{Del67}, along
with very few more recent data from Bates\cite{Dyt90} and Tokyo\cite{Hom88}.
With the recent completion of the electron accelerators at Mainz (MAMI B) and
Bonn (ELSA) and the construction of new spectrometers and detectors it is
now possible
to measure eta photoproduction from threshold at 707 MeV up to 850 MeV at
Mainz and even higher energies at Bonn with a precision similar to the one
obtained in pion photoproduction experiments. A large amount of data has
already been taken and is currently being analyzed\cite{Ant93,Kru93}.

Unlike pion photoproduction, low energy theorems ($LET$) cannot be derived
for eta photoproduction for the following three reasons:
({\it i}) The expansion parameter $\mu=m_{\eta}/m_{N}\approx
0.6$ is too large to provide convergence up to order $\mu^2$; ({\it ii})
due to large $\eta-\eta'$ mixing with a mixing angle of about $20^\circ$ and a
non-conserved axial singlet current $A_0^{\mu}$ for the $\eta'$, there is no
$PCAC$ theorem for eta mesons; ({\it iii}) there are nucleon resonances,
mainly the $S_{11}(1535)$ close at threshold ($W_{thr.}=1486$ MeV)
strongly violating the condition that the internal excitation energy must be
larger than the mass of the meson\cite{Fri83,Dre84}.

Nucleon resonance excitation is the dominant reaction process in
$(\gamma,\eta)$. In contrast to pions which will excite $\Delta(T=3/2)$
as well as $N^*(T=1/2)$ resonances, the $\eta$ meson will only appear in the
decay of  $N^*$ resonances with $T=1/2$. In the low-energy region this is
dominantly the $S_{11}(1535)$ state that decays in 45-55\% into $\eta N$,
the only nucleon resonance with such a strong branching ratio in the $\eta$
channel. This result is even more surprising as a near-by resonance of similar
structure, the $S_{11}(1650)$ has a branching ratio of only 1.5\%.
This "$\eta$ puzzle" is not yet understood in quark models of the nucleon.

Most attempts to describe eta photoproduction on the nucleon\cite{Hic73,Muk91}
have involved Breit--Wigner functions for the resonances and either
phenomenology or a Lagrangian approach to model the background. These
models which contain a large number of free parameters were then adjusted to
reproduce the few available data. In a very different approach,
Ref.~\cite{Ben91} derived a dynamical model which employs
$\pi N \rightarrow \pi N,\pi N \rightarrow \pi\pi N$ and
$\pi^- p \rightarrow \eta n$ to fix the hadronic vertex as well as the
propagators and the $\gamma N \rightarrow \pi N$ to construct the
electromagnetic vertex. This calculation represents a prediction rather
than a fit to the $\gamma N \rightarrow \eta N$ reaction.

Here we extend the model of Ref.~\cite{Ben91} by taking into account the
background  from $s, u$--channel nucleon Born terms and $\rho,\omega$
exchange in the $t$--channel. Since the resonance sector is fixed in our
approach and also the vector meson couplings can be obtained from
independent sources,
we can use this model to extract information on the $\eta NN$
coupling.  Furthermore, we apply the operator to elastic eta photoproduction
on the very light nuclei, $d$, $^3$He, $^3$H and $^4$He. Due to spin and
isospin selection rules, measuring a combination of processes on these light
nuclei with well-known nuclear structure should hopefully allow us a complete
determination of the individual $(\gamma,\eta)$
multipoles for protons and neutrons.

In Sec. II we shortly summarize the resonance model of Ref.~\cite{Ben91}
and describe our full $(\gamma,\eta)$ operator. Sec. III contains our
discussion of the elementary $\eta NN$ vertex.  The possibility to get new
information about nucleon resonances in the eta channel with the help of
polarization observables is discussed in Sec. IV.
The formalism for eta
photoproduction on nuclei is derived in a coupled channel framework in
Sec. V and we present predictions for differential cross sections of
elastic $(\gamma,\eta)$ on $d$, $^3$He, $^3$H and $^4$He. In Sec. VI we
summarize our findings and present a brief outlook.
In the Appendix we present the definitions of the 16 polarization
observables for photoproduction of pseudoscalar mesons and give expansions
in CGLN amplitudes and dominant multipoles.

\section{ETA PHOTOPRODUCTION ON THE NUCLEON}

The dynamical model of Bennhold and Tanabe\cite{Ben91} is based on the
observation that near the $\eta$ production threshold three nucleon resonances
$P_{11}(1440),\,D_{13}(1520)$ and $S_{11}(1535)$ play an important role.
Assuming an isobar model for each partial wave the transition amplitude can
be written as
\begin{eqnarray}
t_{ij}(W)=f_{i}^{\dagger}D^{-1}(W)f_{j}\; ,
\end{eqnarray}
where $W$ is the invariant energy and $i,j=\pi,\eta$ denotes the $\pi N$ and
$\eta N$ channels, respectively. The vertex functions $f_i$ are parametrized
with coupling strengths and formfactors and the $N^*$ propagators are given
by
\begin{eqnarray}
D(W)=W-m_0-\Sigma_{\pi}(W)-\Sigma_{\eta}(W)+\frac{i}{2}\Gamma_{\pi\pi}(W)
\end{eqnarray}
with the bare resonance mass $m_0$.

The self-energy $\Sigma$ associated with the $\pi N$ and $\eta N$
intermediate states is given by
\begin{eqnarray}
\Sigma_{i}(W)=\int_0^{\infty}\frac{q^2dq}{(2\pi)^3}\,\frac{M}{2w_i(q)E_N(q)}
\left(\frac{q}{m_i}\right)^{2l}\frac{g_i^2(1+q^2/\Lambda_i^2)^{-2-l}}
{W-w_i(q)-E_N(q)+i\epsilon}
\end{eqnarray}
with $w_i(q)=\sqrt{m_i^2+q^2},\,E_N(q)=\sqrt{M_{}^2+q^2}$ and $M$ denoting
the nucleon mass. The two-pion decay width $\Gamma_{\pi\pi}$ is parametrized
with one free parameter. The six parameters in this approach have been
determined  for each partial wave by a least-squares fit to all data of the
reactions  $\pi N \rightarrow \pi N,\,\pi N \rightarrow \pi\pi N$ and
$\pi^- p \rightarrow \eta n$ and can be found in Ref.\cite{Ben91}.

In order to use this operator more conveniently, especially in nuclear
applications with multidimensional integrals, we have obtained simple
parametrizations of the self-energy $\Sigma$, Eq.~(3), in very good agreement
with the exact numerical values.
\begin{mathletters}
\begin{eqnarray}
Re\,\Sigma&=&a + (b_1\sqrt{x}+b_2 x^2)\Theta(-x) + (c_1 x+c_2
x^2)\Theta(x)\,,\\
I\!m\,\Sigma&=&(d_1\sqrt{x}+d_2 x+d_3 x^2)\Theta(x)
\end{eqnarray}
\end{mathletters}
with $x=(W-M-m_i)/m_{\pi}$, $i=\pi,\eta$ and the step function
$\Theta(x)$. The parameters are given in Table I.
Finally the decay width in the 2$\pi$--channel is given by
\begin{eqnarray}
\Gamma_{\pi\pi}(W)=\gamma\, x\, \Theta(x)\,,\qquad x=(W-M-2m_{\pi})/m_{\pi}
\end{eqnarray}
with $\gamma(S_{11})=4.3 MeV$, $\gamma(P_{11})=80.3 MeV$ and
$\gamma(D_{13})=24.2 MeV$.

With the hadronic vertices and propagators determined, the
photoproduction amplitudes for $(\gamma,\pi)$ and $(\gamma,\eta)$
are given by
\begin{eqnarray}
t_{i\gamma}(W)\,=\,V_{i\gamma}^B(W)\,+\,f_{i}^{\dagger}D^{-1}(W)
\tilde{f_{\gamma}}\,,
\end{eqnarray}
where $V_{i\gamma}^B$ are the Born terms and $\tilde{f_{\gamma}}$ is the
electromagnetic vertex. The latter was determined by using pion
photoproduction data. By this way no free parameters are introduced in the
$(\gamma,\eta)$ process. Since Ref.\cite{Ben91} neglected the Born terms
in the $\eta$--channel, $V_{\eta\gamma}^B\equiv0$, the model
consisted of four $(\gamma,\eta)$ multipoles only, the $S_{11}(1535)$
appears in the dominant $E_{0+}$, the $P_{11}(1440)$ in the $M_{1-}$ and
the $D_{13}(1520)$ in the $E_{2-}$ and $M_{2-}$.

While neglecting the $(\gamma,\eta)$ Born terms was within the
uncertainties of the older experimental data for the proton,
they play a more important role when comparison with better data becomes
possible.  Furthermore, including the background properly
becomes necessary in nuclear reactions like the coherent $\eta$
photoproduction on $^4$He, where the dominant excitation of the $S_{11}$
resonance is forbidden.

The evaluation of the background terms is
straightforward and in complete analogy to $(\gamma,\pi^0)$ except for the
fact that the $\eta$ is an isoscalar meson and two types of $\eta NN$
couplings are possible: Pseudovector (PV) and pseudoscalar (PS). The latter
one is not ruled out by LET as in the case of $(\gamma,\pi)$.

The effective Lagrangians for both types of the $\eta NN$ coupling are given by
\begin{eqnarray}
{\cal L}_{\eta NN}^{PS}=-ig_{\eta}\bar{\psi}\gamma_5\psi\phi_{\eta}\,,
\qquad
{\cal L}_{\eta NN}^{PV}=\frac{g_{\eta}}{2M}\bar{\psi}\gamma_{\mu}
\gamma_5\psi\partial^{\mu}\phi_{\eta}\,.
\end{eqnarray}
With the electromagnetic Lagrangian
\begin{eqnarray}
{\cal L}_{\gamma NN}^{PS}=-e\bar{\psi}\gamma_{\mu}\frac{1+\tau_0}{2}
\psi A^{\mu}+\frac{e}{4M}\bar{\psi}(\kappa^S+\kappa^V\tau_0)\sigma_{\mu\nu}
\psi\,F^{\mu\nu}\,,
\end{eqnarray}
where $\kappa^S=-0.06$ and $\kappa^V=1.85$ are the isoscalar and isovector
anomalous magnetic moments and
$F^{\mu\nu}=\partial^{\nu}A^{\mu}-\partial^{\mu}A^{\nu}$
we can evaluate the $s$-- and $u$--channel Born terms.
Expressed in the $CGLN$ basis
\begin{eqnarray}
F=iF_{1}\,\vec{\sigma}\cdot\vec{\epsilon}+F_{2}\,\vec{\sigma}\cdot\hat{\vec q}
\,\sigma\cdot(\hat{\vec{k}}\times\vec{\epsilon})+
iF_{3}\,\vec{\sigma}\cdot\hat{\vec k}\,\hat{\vec q}\cdot\vec{\epsilon}+
iF_{4}\,\vec{\sigma}\cdot\hat{\vec q}\,\hat{\vec{q}}\cdot\vec{\epsilon}
\end{eqnarray}
we obtain the following amplitudes for pseudoscalar coupling
\begin{mathletters}
\begin{eqnarray}
F_1(PS)&=&g_{\eta}\,C\left[\left(-e_N+\frac{W-M}{2M}\kappa_N\right)
D+\frac{(t-m_{\eta}^2)\kappa_N}{2M(W-M)(u-M^2)}\right]\,,\\
F_2(PS)&=&g_{\eta}\,\frac{C\mid \vec{q} \mid}{E_2+M}\left[\left(e_N+
\frac{W+M}{2M}\kappa_N\right)D+
\frac{(t-m_{\eta}^2)\kappa_N}{2M(W+M)(u-M^2)}\right]\,,\\
F_3(PS)&=&g_{\eta}\,C\mid \vec{q} \mid \left[2e_N\frac{W-M}{t-m_{\eta}^2}
D-\frac{\kappa_N}{M(u-M^2)}\right]\,,\\
F_4(PS)&=&g_{\eta}\,\frac{C\mid \vec{q} \mid^2}{E_2+M^2}
\left[-2e_N\frac{W+M}{t-m_{\eta}^2}D-
\frac{\kappa_N}{M(u-M^2)}\right]\,,
\end{eqnarray}
\end{mathletters}
where
$t=2(\vec{k}\cdot\vec{q}-E_{\gamma}E_{\pi})+m_{\eta}^2\,$,
$u=-2(\vec{k}\cdot\vec{q}+E_{\gamma}E_2)+M^2\,$,
$E_{1(2)}$ is the nucleon
energy in the initial (final) state,
and
\begin{eqnarray}
C=-e\,\frac{W-M}{8\pi W}\sqrt{(E_1+M)(E_2+M)}\,,\qquad
D=\frac{1}{W^2-M^2}+\frac{1}{u-M^2}\,.
\end{eqnarray}
For pseudovector coupling we get
\begin{eqnarray}
F_1(PV)=F_1(PS)-g_{\eta}\frac{C\kappa_N}{2M^2},
\qquad
F_2(PV)=F_2(PS)+g_{\eta}\frac{C\mid \vec{q} \mid
\kappa_N}{2M^2(E_2+M)},
\end{eqnarray}
and no change for $F_3$ and $F_4$. Note that the difference between
pseudoscalar and pseudovector coupling arises only from the anomalous
magnetic moment of the nucleon.

In the above equations (where $N=p,n$) the amplitudes are expressed for the
protons and neutrons separately with $e_p=1$, $e_n=0$ and $\kappa_p=1.79$,
$\kappa_n=-1.91$. Alternatively we can define the isoscalar and isovector
amplitudes $F_i^{(0)}$ and $F_i^{(1)}$ via
\begin{eqnarray}
F_i=F_i^{(0)}+F_i^{(1)}\tau_0
\end{eqnarray}

Due to the decay of the vector mesons ${\mathrm V}(J^{\pi};T)=\omega(1^-;0)$
and
$\rho(1^-;1)$ into $\eta\,\gamma$ we also have to include the $t$--channel
Born diagrams which we evaluate from the Lagrangians
\begin{eqnarray}
{\cal L}_{{\mathrm V}NN}=-g_V\bar{\psi}\gamma_{\mu}\psi V^{\mu}+
\frac{g_T}{4M}\bar{\psi}\sigma_{\mu\nu}\psi\,V^{\mu\nu}\,,
\qquad
{\cal L}_{{\mathrm V}\eta\gamma}=\frac{e\lambda_{\mathrm V}}{4m_{\eta}}
\varepsilon_{\mu\nu\lambda\sigma}F^{\mu\nu}V^{\lambda\sigma}\phi_{\eta}
\end{eqnarray}
with
$V^{\mu\nu}=\partial^{\nu}V^{\mu}-\partial^{\mu}V^{\nu}$
like the electromagnetic field tensor $F^{\mu\nu}$. This yields to the $CGLN$
amplitudes
\begin{mathletters}
\begin{eqnarray}
F_1(\mathrm V)&=&\frac{\lambda_{\mathrm V}\,C}{m_{\eta}(t-m_{\mathrm V}^2)}
\left[-\frac{g_T}{2M}t+\left(\frac{t-m_{\eta}^2}{2W-2M}+W-M\right)
g_V\right]\,,\\
F_2(\mathrm V)&=&\frac{\lambda_{\mathrm V}\,C}{m_{\eta}(t-m_{\mathrm V}^2)}\,
\frac{\mid \vec{q} \mid}{E_2+M}
\left[\frac{g_T}{2M}t+\left(\frac{t-m_{\eta}^2}{2W+2M}+W+M\right)
g_V\right]\,,\\
F_3(\mathrm V)&=&\frac{\lambda_{\mathrm V}\,C}{m_{\eta}(t-m_{\mathrm V}^2)}
\mid \vec{q} \mid \left[\frac{g_T}{2M}(W-M)-g_V\right]\,,\\
F_4(\mathrm V)&=&-\frac{\lambda_{\mathrm V}\,C}{m_{\eta}(t-m_{\mathrm V}^2)}\,
\frac{\mid \vec{q} \mid^2}{E_2+M}
\left[\frac{g_T}{2M}(W+M)+g_V\right]\,.
\end{eqnarray}
\end{mathletters}
Due to the isospin, the $\omega$ contributes only to $F_i^{(0)}$ and $\rho$
only to $F_i^{(1)}$.

In Table II we give the coupling constants and cut--off masses for the
background contributions. For the vector mesons we have introduced dipole
formfactors
$F(\vec{q}^2)=(\Lambda_{\mathrm V}^2-m_{\mathrm V}^2)^2/(\Lambda_{\mathrm V}^2+
\vec{q}^2)^2$ at the ${\mathrm V} NN$ vertex given by the Bonn
potential\cite{Mach89}, for
the $\eta NN$ coupling the formfactors turned out to be insensitive
in the energy region of our interest and have
been ignored. The main effect would result in a renormalization of the
coupling constant.
The electromagnetic V$\eta\gamma$ couplings are obtained from
the partial decay widths of the vector mesons.

\section{THE $\eta NN$ COUPLING}

In contrast to the $\pi N$-interaction, little is known about the $\eta
N$-interaction and, consequently, about the $\eta NN$ vertex. As it was
mentioned before, in the
case of pion scattering and pion photoproduction the $\pi NN$ coupling
is preferred to be pseudovector (PV), in accord with current algebra
results and chiral symmetry.  However, because the eta mass is so much
larger than the pion mass - leading to large SU(3) x SU(3) symmetry
breaking - and because of the $\eta-\eta'$ mixing there is no compelling reason
 to select the PV rather than
the PS form for the $\eta NN$ vertex.
%We therefore explore both
%possibilities in confidence that the new experimental data from Mainz
%and Bonn will be able to distinguish between the two different forms.

The uncertainty regarding the structure of the $\eta NN$ vertex
extends to the magnitude of the coupling constant.
This coupling constant $g_{\eta NN}^2/4\pi$ varies between $0$ and $7$
with the large couplings arising from fits of one boson exchange potentials.
Typical values obtained in
fits with the Bonn potential\cite{Bro90} can lie anywhere between 3 - 7.
However, including the $\eta$ yields only small effects in fitting the
$NN$ phase
shifts and, furthermore, provides an insignificant contribution to nuclear
binding at normal nuclear densities.
{}From SU(3) flavor symmetry all coupling constants between the meson octet
and the baryon octet are determined by one free parameter $\alpha$, giving
\begin{equation}
\frac{g_{\eta NN}^2}{4 \pi} = \frac{1}{3}(3-4\alpha)^2\;
\frac{g_{\pi NN}^2}{4 \pi}\;,
\end{equation}
resulting in values for the coupling constant
 between $0.8$ and $1.9$ for commonly used values of
$\alpha$ between $0.6 - 0.65$, depending on the F and D strengths chosen as the
 two types of
SU(3) octet meson-baryon couplings.
Other determinations of the $\eta NN$
coupling employ reactions involving the eta, such as
$\pi^- p \rightarrow \eta n$, and range from 0.6 - 1.7\cite{Pen87}.
Smaller values are supported by $NN$ forward dispersion
relations\cite{Grein80} with
$g_{\eta NN}^2/{4 \pi}+g_{\eta' NN}^2/{4 \pi}
\leq 1.0$.
There is some rather indirect evidence that also favors a small value for
$g_{\eta NN}$.
In Ref.\cite{Pie93}, Piekarewicz calculated the $\pi$--$\eta$
mixing amplitude in the hadronic model where the mixing was generated by
$\bar N N$ loops and thus driven by the proton--neutron mass difference.
To be in agreement with results from chiral perturbation theory the $\eta
NN$ coupling had to be constrained to the range 0.32 -- 0.53.  In a very
different approach, Hatsuda\cite{Hat90} evaluated the proton matrix
element of the flavor singlet axial current in the large $N_C$ chiral
dynamics with an effective Lagrangian that included the $U_A$(1)
anomaly.  In this framework, the EMC data on the polarized proton
structure function (which have been used to determine the "strangeness
content" of the proton) can be related to the $\eta 'NN$ and the $\eta
NN$ coupling constants.  Again, his analysis prefers small values for
both coupling constants.  Nevertheless, from the above
discussion it seems clear that the $\eta NN$ coupling constant is much
smaller compared to the corresponding $\pi NN$ value of around 14.

Since in our model the resonance sector is well constrained by other
related but independent reactions we use the $(\gamma,\eta)$ data to
extract information on the $\eta NN$  vertex.
In Table III we give the individual contributions to the threshold $E_{0+}$
multipole for protons and neutrons. While the $S_{11}$ contribution is
complex, the background contribution from Born terms and vector mesons is
real. In the table, the Born PS and PV terms are calculated for a coupling
of $g_{\eta NN}^2/4\pi = 1$ and scales proportional to $g_{\eta NN}$. Note
that for a coupling of $g_{\eta NN}^2/4\pi = 0.33$ the background vanishes
in a pseudoscalar model.
%Such a coupling strength seems to be favoured by
%preliminary results of the Mainz experiment\cite{Kru93}.

In Fig. 1, we show the
sensitivity of the total cross section close to threshold when varying
the coupling constant from 0 to 3 for the PS and from 0 to 10 for
the PV form.  There is
a large variation of more than a factor of two at 750 MeV for the PS
case while changing the value with PV structure modifies the cross
section only by a relatively small amount.  The difference is due to the
fact that the PV vertex contains momentum dependence which influences
mostly the p-wave multipoles.  On the other hand, at threshold
the total cross section is dominated by the s-wave multipole due to the
$S_{11}(1535)$ resonance.  A similar effect can be observed in
$(\gamma,\pi^\circ)$ at threshold where the PS Born terms overpredict the
LET prediction by a large amount while the PV form agrees with the small
LET value.

However, using only the total cross section data does not allow to
uniquely determine the coupling constant.
As shown in Fig. 2, one obtains similar total cross sections for a PS coupling
of 0.1 and a PV coupling of 6.0.  Data that would
fall below this curve could only be explained with a pseudoscalar
model. For example, a coupling strength of $g_{\eta NN}^2/4\pi=0.4$
gives results very close to the pure resonance contribution.
Large PS couplings around 1.0 or 1.4 suggested in previous
eta photoproduction studies
\cite{Muk91} would be consistent only with data considerably below 15
$\mu b$ at the maximum. Using only the old data shown in Fig. 2 no
definite conclusion can be reached at this point.

Since the total cross section alone cannot unambiguously delineate
between the two different coupling modes, we present in Fig. 3
calculations for the differential
cross section at the four different photon energies
that were used in the Mainz experiment currently under analysis.
 Note that both computations performed in the
PS- and PV-model give roughly the same total cross section. There is a
clear distinction in the forward-backward asymmetry of the angular
distribution between  the PS- and PV-model. While the older data at 750 MeV
(shown in Fig. 3)
cannot uniquely distinguish between the two coupling schemes the new
preliminary Mainz data with very small error bars indicate a clear
preference for a PS-vertex with a small coupling constant. The variation in
the angular distributions is again due to the $p$-wave multipoles. In
particular, the $M_{1-}$ multipole changes sign between PS and PV coupling.

\section{MULTIPOLES AND POLARIZATION OBSERVABLES}

In order to obtain a detailed understanding of the eta photoproduction
process and resonance phenomena associated with it, it would be
desirable to perform
a multipole analysis along the lines that have been pursued in pion
photoproduction for over twenty years.
Multipoles offer the possibility to especially
study  resonance properties in detail since - due to their particular
quantum numbers - resonances contribute only to one specific multipole
for $J=1/2$ and to two multipoles (electric and magnetic) for $J\geq 3/2$.
In contrast to pion photoproduction,
where the $\Delta_{33}(1232)$ resonance dominates
for the first 400 MeV above
threshold, there are
three resonances in eta photoproduction right at threshold - the dominant
$S_{11}(1535)$, and the weaker $D_{13}(1520)$ and $P_{11}(1440)$ states.
Additional resonances - such as the $P_{11}(1710)$ - are expected
to contribute significantly in the region
of 200-400 MeV above threshold.
Furthermore, the suppression of the Born terms in the $(\gamma,\eta)$
process due to the small $\eta NN$ coupling constant offers the opportunity
to extract valuable resonance information from the $(\gamma,\eta)$
multipoles.

In Fig. 4 we present the real and imaginary part of the $E_{0+}$ multipole
as a function of the photon energy. Clearly, the $S_{11}$ resonance
dominates this multipole and provides the only contribution to the
$Im(E_{0+})$ since the Born terms  and vector meson contributions -
being small and opposite in sign - have no imaginary part. The magnitude of
the $S_{11}$ resonance at threshold shows the futility of extracting
$LET$ values at threshold. Regarding the $\eta NN$ vertex, the $E_{0+}$ is
insensitive to the difference between the PS- and PV- coupling by properly
adjusting the coupling constants as was shown in Fig. 2.

This situation changes dramatically  for the real part of the
$M_{1-}$ multipole shown in Fig. 5. This $p$-wave multipole which is
positive for PS- but negative for PV-coupling is responsible for the
variation in the
forward-backward asymmetry in the differential cross sections of Fig. 3.

In Fig. 6 we present the recoil polarization $P$ which is
proportional to the $M_{1-}$ multipole
\begin{equation}
P=-2 sin(\theta) \frac{q}{k \sigma(\theta)} Im(E_{0+}^*\,M_{1-}\,
+ \dots)\;.
\end{equation}
Thus, the PS-coupling leads to a positive $P$, supported by a single data
point available in this energy region, while PV-coupling leads a negative
recoil polarization. Besides the sensitivity to the nature of the
$\eta NN$-vertex, Fig. 6 also shows the presence of the $P_{11}(1440)$
state (Roper resonance) in the real part of the $M_{1-}$ multipole.
Just as this resonance is not easily identified in other electromagnetic
reactions it is not very noticeable in the eta photoproduction process
as well.

In Fig. 7 we show the effect of the $D_{13}(1520)$ and the $P_{11}(1440)$
resonances in the differential cross section, $d\sigma/d\Omega$, the
three single-polarization observables $\Sigma,T,P$ and the four
double-polarization observables $E,F,G,H$ that require polarization
of the beam and the target simultaneously.
The energy of 752 MeV is chosen to be in the region where the $D_{13}$
resonance has its maximum contribution. Omitting the $D_{13}$ from our
calculations gives dramatic effects in the beam asymmetry $\Sigma$ as well
as in the double-polarization observable $G$, but $T,P,F$ and $H$
also exhibit a significant sensitivity to the $D_{13}$ state.
This is in contrast to the differential cross
section that shows very little sensitivity to the $D_{13}$ state.
This behavior can
be understood in terms of the multipole contributions.
Especially the photon asymmetry $\Sigma$
, which changes from its maximum value around 0.4 at
$\theta=90^0$ to almost zero when the $D_{13}(1520)$ is not included,
the expansion into leading multipoles yields
\begin{equation}
\Sigma =3 sin^2(\theta) \frac{q}{k \sigma(\theta)}
Re[E_{0+}^*\,(E_{2-}+M_{2-}) + \dots]\;.
\end{equation}
As in the previous case with the $P_{11}$, the interference of the D-wave
multipoles with the dominant $E_{0+}$ gives such an enhanced sensitivity.
While in general all observables of Fig. 7 are different, in this situation
with $S_{11}$ dominance (between 700 and 900 MeV) we find similar
structures of $\Sigma$ and $G$, $T$ and $F$, and $P$ and $H$.
In future experiments with polarized photon beams and polarized targets
comparisons of those pairs of observables can give valuable information
on small background multipoles. Furthermore, it can be used to
separate real and imaginary parts of the resonance multipoles.
Currently, such polarization experiments are already in preparation
at LEGS in Brookhaven, GRAAL in Grenoble and CEBAF with an expected start
at the end of 1995.

As an example for the $l=2$ multipoles, Fig. 8 depicts the real part of
the $E_{2-}$ multipole. Note that this multipole is again clearly dominated
by its resonance contribution since the Born terms and vector mesons almost
cancel. Therefore, using total and differential cross section measurements
should reveal the strong $E_{0+}$ multipole, while the smaller $l=1$ and
$l=2$ multipoles could be extracted with the help of polarization
experiments.

In the appendix we give the complete structure and definitions of the 16
observables in photoproduction of pseudoscalar mesons as well as the
expansion into leading multipoles.

\section{ETA PHOTOPRODUCTION ON NUCLEI}

Eta photoproduction on nuclei can be developed in a straightforward way
by the same method which has been applied very successfully in pion
photoproduction\cite{Kam91}. In momentum space the nuclear photoproduction
amplitude can be written as
\begin{equation}
F_{\eta \gamma}(\vec{q},\vec{k}) = V_{\eta \gamma}(\vec{q},\vec{k}) -
\frac{a}{(2\pi)^2} \sum_{{i=\pi,\eta}} \int \frac{d^3q'}{M_i(q')} \frac{F_{\eta
i}(\vec{q},\vec{q}\,')\,V_{i\gamma}(\vec{q}\,', \vec{k})}
{W_{\eta}(q) -W_i(q') + i \epsilon} \, \, ,
\end{equation}
where $\vec{k}$ is the photon, and $\vec{q}$ is the eta or pion momentum. The
total energy in the $\eta$-nucleus and $\pi$-nucleus channels is denoted by
$W_i(q) = E_i(q) + E_A(q)$, the reduced mass is given by
$M_i(q) = E_i(q) E_A(q)/W_i(q)$ and $a=(A-1)/A$.

$V_{\eta \gamma}$ is expressed in terms of the free eta--nucleon
photoproduction t--matrix
\begin{eqnarray}
V_{\eta \gamma}(\vec{q},\vec{k})=
-\frac{\sqrt{M_{\eta}(q)M_{\gamma}(k)}}{2\pi}<\eta(\vec{q}),f \mid
\sum^A_{j=1} \hat{t}_{\gamma N}(j)\mid i,\gamma (\vec{k})> ,
\end{eqnarray}
where $\mid i>$ and $\mid f>$ denote the nuclear initial and final
states, respectively, and  $j$ refers to the individual target nucleons.

   Using the KMT version of multiple scattering theory\cite{Ker59} the meson
scattering amplitude $F_{ij}$ is constructed as a solution of the
Lippmann-Schwinger equation
\begin{equation}
F_{ij}(\vec{q}\,',\vec{q}) =V_{ij}(\vec{q}\,',\vec{q})-
\frac{a}{(2\pi)^2} \sum_{l=\pi,\eta} \int \frac{d^3q''}{M_l(q'')}
\frac{V_{il}(\vec{q}\,',\vec{q}\,'')\,F_{lj}(\vec{q}\,'',\vec{q})}
{W_j(q) -W_l(q') + i \epsilon} \, \, ,
\end{equation}
Here  the meson-nuclear interaction is described by the first-oder potential
$V_{ij}=(V_{\pi\pi},\,V_{\eta\pi},\,V_{\eta\eta})$ which is related to the
corresponding free $t_{ij}$ matrix of meson-nucleon interaction\cite{Kam93}.

At present our calculations have been carried out only for the first part
of Eq.~(19), the plane wave impulse approximation ($PWIA$). At this level,
however, we do not perform any approximation treating the full spin degrees
of freedom and taking Fermi motion effects of the nucleon into account by
performing the integration in momentum space. For the deuteron\cite{Lac81} and
$^3$He/$^3$H\cite{Bra75} we use realistic nuclear wave functions.
In the case of $^4$He with $J=T=0$ a phenomenological nuclear formfactor
is used which has been extracted from the charge distribution of $^4$He.

By studying eta photoproduction on light nuclei with well--known
nuclear structure we can learn about details of the elementary production
operator which are difficult to see in the elementary reaction or, as for the
neutron amplitude, are not experimentally accessible. In the deuteron case
only the isoscalar amplitude contributes; in $^3$He the two protons saturate
to spin 0 and contribute only to a very small part via the non--spin
amplitude  from the $P_{11}(1440)$ and background terms, while the residual
neutron gives rise to a strong $E_{0+}$ amplitude. Finally, in the case of
$^4$He we can study the coherent amplitude of $t_{\eta\gamma}$ which is the
isoscalar non--spin flip part and arises from small magnetic
multipoles, e.g. $M_{1-}$ and $M_{1+}$.

Besides studying the elementary amplitude, eta photoproduction offers the
possibility to learn more about the $\eta N$- and $\eta A$-interaction.
Recently it has been suggested by Wilkin\cite{Wil93} that the
very large production
cross sections found in the $pd\rightarrow \eta ^3He$ reaction near threshold
could be due to an $\eta N$ scattering length that is much
larger than the value extracted from  a coupled-channels analysis of the
$\pi^-p\rightarrow\eta n$ and $\pi N\rightarrow\pi N$ data\cite{Bha85}.
Based on the
K-matrix formalism, Wilkin was able to reproduce the strong threshold
energy dependence of $\pi d \rightarrow \eta ^3He$ by assuming an $\eta N$
scattering length with a real part more than twice as large as in previous
analyses. Should this conclusion turn out to be true, it would have
dramatic implications for the $\eta$-nucleus interaction. Most importantly,
a larger value for the scattering length
might indicate a larger probability for the
presence of a "bound" $\eta$-state for lighter nuclei such as $^3$He than
had been expected. In fact, this seems to be required to explain the
cusp-like structure seen for near-threshold production in the
$\pi d\rightarrow {^3He} X$ reaction for a missing mass close to the
$\eta$-mass. An uncertainty in Wilkin's analysis arises from a cross
section measurement of the $p d \rightarrow \eta ^3He$ process very close to
threshold ($\sim$200 keV) that seems to contradict his description of the
energy dependence. However, if this one point - which suffers from large
systematic uncertainties - is ignored, impressive agreement with the
experiment is achieved. A new experiment at Saclay has already been
approved to explore the energy dependence  in the region very close to
threshold
\cite{Bris94}. This effect may also remove the discrepancy between experiment
and theory for the reaction $^3He(\pi^-,\eta)^3H$\cite{Kam93}.
In a future study we
will address the significance of this large $\eta^3He$-interaction in the
$^3He(\gamma,\eta)^3He$ cross section calculated in a full DWIA framework.

In Fig. 9 we show the differential cross sections for all light nuclei
up to $^4$He. Whereas the angular distribution is rather flat for nucleons,
as shown before, it appears more and more peaked in forward
direction for $A>1$. This reflects the signature of the nuclear formfactors
as the momentum transfer in $\eta$ photoproduction is rather large,
$Q^2$ = 7.8 $fm^{-2}$ at threshold. The biggest cross section can be expected
for the trinucleon; it is proportional to the free nucleon cross section
multiplied by the square of the trinucleon formfactor.
The triton cross section is about a factor of two larger than that for
$^3$He since the spin-flip amplitude on the proton  - which is the largest
among the possible spin-flip and non-spin-flip amplitudes on the proton and the
neutron - cannot contribute due to the Pauli principle. Therefore, the
single proton in $^3$H provides more strength than the two protons
(coupled mostly to spin 0) in $^3$He.

Around $90^\circ$ the cross section on the deuteron
gains over the trinucleon, even though it is in large disagreement with
the few available data. This may be due to the rather small isoscalar
amplitude predicted by our model,
 $E_{0+}^{(0)}/E_{0+}^{(p)}=0.15$.
In the naive quark model the electromagnetic excitation of the $S_{11}$
is almost entirely isovector. However, Rosenthal, Forest and Gonzales
\cite{Ros91} have shown that a color--hyperfine interaction, responsible
also for the E2/M1 ratio of the $\Delta$ excitation, could enhance the
isoscalar $S_{11}$ excitation considerably.

Only an unrealistically large isoscalar amplitude of
$E_{0+}^{(0)}/E_{0+}^{(p)}=0.8$ can explain the deuteron data in $PWIA$.
As it has been shown in Ref.~\cite{Hal89} if the final state interaction and
the coupled $\pi\eta$--channels are taken into account then the discrepancy
can be explained with $E_{0+}^{(0)}/E_{0+}^{(p)}=0.6$. It also remains to be
seen if the isoscalar amplitude is really as small as all present
models predict. Finally, the coherent cross section for $^4$He vanishes for
$\theta=0$ and reaches roughly the $10 nb$ level in a small
angular region. For most angles it falls below $1 nb$. It is an
experimental challenge to measure this reaction which provides a clean
observation of the background multipoles and of the Roper resonance, as
the dominant $S_{11}$ resonance is suppressed by spin and isospin.

\section{SUMMARY AND CONCLUSIONS}

We have presented a model for eta photoproduction on the nucleon
that includes nucleon Born terms and t-channel vector meson exchanges in
addition to the nucleon resonances $S_{11}(1535)$, $P_{11}(1440)$ and
$D_{13}(1520)$.  The resonance sector is fixed by using data from other
hadronic and electromagnetic reactions such as pion scattering and
photoproduction and pion induced eta production.  Vector meson couplings
are determined from their radiative decay widths and the
$NN$--interaction.  This allows using the new experimental data from
Bonn and Mainz to extract information on the $\eta NN$ coupling.  While
the total $(\gamma,\eta)$ cross section on the proton can be well
reproduced by either a small coupling constant with PS--coupling or a
large value with PV--form, the angular distribution singles out the
small constant.

While the resonance parameters of the $S_{11}$ can be extracted from
high-precision total cross section data, the smaller resonances are hidden both
in the differential and total cross sections. Here, polarization
observables will provide a powerful tool to constrain
these small resonance couplings,
 as demonstrated for the case of the $D_{13}(1520)$.
Such experiments have been proposed and will soon be performed at the LEGS
facility at Brookhaven National Laboratory and at the new facility GRAAL
at Grenoble.

Finally, we have applied our operator to eta photoproduction on
light nuclei d, $^3$He, $^3$H and $^4$He.
Experiments on these nuclei are necessary to obtain complete
information on the isospin structure of the $(\gamma,\eta)$ amplitude.
At forward angles the cross sections for $(\gamma,\eta)$ on the trinucleon
are reasonably large and should be measurable. Here especially the threshold
region would be interesting where one could learn more about the
eta-nucleus interaction.

\acknowledgments

We would like to thank Gisela Anton, Hans Str\"oher, Bernd Krusche and
Michael Wilhelm for fruitful discussions.
This work was supported by the Deutsche Forschungsgemeinschaft (SFB201),
the U.S. DOE grant DE-FG05-86-ER40270 and the Heisenberg-Landau program.
One of the authors, S.K. wants to thank the theory group of Prof. Drechsel and
Monika Baumbusch for the very kind hospitality at the Institut f\"ur
Kernphysik in Mainz.

\appendix
\section{Polarization Observables}
Following the notation of Barker et al.\cite{Bar75} we can write the
differential cross section
\begin{description}
\item[a)] for beam and target polarization
\begin{eqnarray}
\frac{d\sigma}{d\Omega} & = & \sigma_0
\{1-P_T\Sigma\cos 2\varphi\\
&+& P_x(-P_T H\sin 2\varphi + P_\odot F)-P_y(-T+P_T P\cos 2\varphi)\nonumber\\
&-& P_z(-P_T G\sin 2\varphi + P_\odot E)\}\;,\nonumber
\end{eqnarray}
\item[b)] for beam and recoil polarization
\begin{eqnarray}
\frac{d\sigma}{d\Omega} & = & \sigma_0
\{1-P_T\Sigma\cos 2\varphi\\
&+& P_{x'}(-P_T O_{x'}\sin 2\varphi - P_\odot C_{x'})
-P_{y'}(-P+P_T T\cos 2\varphi)\nonumber\\
&-& P_{z'}(P_T O_{z'}\sin 2\varphi + P_\odot C_{z'})\}\;,\nonumber
\end{eqnarray}
\item[c)] for target and recoil polarization
\begin{eqnarray}
\frac{d\sigma}{d\Omega} & = & \sigma_0
\{1+P_{y'}P + P_{x}(P_{x'}T_{x'}+P_{z'}T_{z'})\\
&+&P_{y}(T+P_{y'}\Sigma)-P_{z}(P_{x'}L_{x'}-P_{z'}L_{z'})\}\;,\nonumber
\end{eqnarray}

\end{description}
where $P_T$ and $P_\odot$ denote the degree of linear and right-handed
circular photon polarization.
$(P_x,P_y,P_z)$ is the polarization of the target in the right-handed
frame $\{x,y,z\}$, with $\hat{z}$ along the photon axis and
$\hat{y}=\hat{k_\gamma}\times\hat{q_\pi}/\sin \theta$.
The spin of the recoil nucleon is analyzed as $(P_{x'},P_{y'},P_{z'})$
in the frame $\{x',y',z'\}$ with $\hat{z'}$ along the meson axis and
$\hat{y'}=\hat{y}$. The angle $\theta$ of the meson as well as all
other quantities are measured in the $cm$ frame. The azimuthal angle
$\varphi$ of the vector of linear photon polarization is measured
counter-clockwise from the scattering plane, e.g. $\varphi=\pi/2$ for
a photon polarization $\hat{\epsilon_\bot}$ along the $\hat{y}$-axis.
The unpolarized cross section $\sigma_0$ will be expressed in terms
of the transverse response function R$_T$,
\begin{equation}
\sigma_0 \equiv \left.\frac{d\sigma}{d\Omega}\right|_{unpolarized} =
\frac{q}{k} R_T\;.
\end{equation}

\section{CGLN amplitudes}
The 16 polarization observables for pseudoscalar meson photoproduction
can be expressed in terms of the four complex CGLN amplitudes of
Eq. (9):

\begin{eqnarray*}
R_T &=& Re\{\mid F_1\mid ^2 + \mid F_2\mid ^2 - 2\cos\theta F_1^*F_2 +
\nonumber\\ &+&
\frac{\sin^2\theta}{2} (\mid F_3\mid^2 + \mid F_4\mid^2 +
2F_2^*F_3 + 2F_1^*F_4 + 2\cos\theta F_3^*F_4)\}
\nonumber\;,\\
R_T \Sigma &=& -\frac{\sin^2\theta}{2}\,Re\{\mid F_3\mid ^2 + \mid F_4\mid ^2
 + 2 (F_2^*F_3 + F_1^*F_4 + \cos\theta F_3^*F_4)\}
\nonumber\;,\\
R_T T &=& \sin\theta \,Im \{F_1^*F_3 - F_2^*F_4 + \cos{\theta}( F_1^*F_4 -
F_2^*F_3) - \sin^2\theta F_3^*F_4\}
\nonumber\;,\\
R_T P &=& \sin\theta \,Im \{F_2^*F_4 - 2F_1^*F_2 - F_1^*F_3 +
\cos{\theta}( F_2^*F_3 - F_1^*F_4) + \sin^2\theta F_3^*F_4\}
\nonumber\;,\\
R_T G &=& \sin^2\theta\,Im\{F_2^*F_3 + F_1^*F_4\}
\nonumber\;,\\
R_T H &=& \sin\theta \,Im\{2F_1^*F_2 + F_1^*F_3 - F_2^*F_4 -
\cos{\theta}( F_2^*F_3 - F_1^*F_4)\}
\nonumber\;,\\
R_T E &=& Re\{\mid F_1\mid^2 + \mid F_2\mid ^2 - 2\cos\theta F_1^*F_2 +
\sin^2\theta(F_2^*F_3 + F_1^*F_4)\}
\nonumber\;,\\
R_T F &=& \sin\theta\,Re\{F_1^*F_3 - F_2^*F_4 + \cos\theta(F_1^*F_4-F_2^*F_3)\}
\nonumber\;,\\
R_T O_{x'} &=& -\sin\theta \,Im\{F_1^*F_4 - F_2^*F_3 +
\cos{\theta}( F_1^*F_3 - F_2^*F_4)\}
\nonumber\;,\\
R_T O_{z'} &=& -\sin^2\theta \,Im \{F_1^*F_3 + F_2^*F_4\}
\nonumber\;,\\
R_T C_{x'} &=& \sin\theta\,Re \{\mid F_1\mid^2-\mid F_2\mid^2 + F_1^*F_4 -
 F_2^*F_3 + \cos\theta(F_1^*F_3 - F_2^*F_4)\}
\nonumber\;,\\
R_T C_{z'} &=&  Re \{2F_1^*F_2 + \sin^2\theta(F_1^*F_3+F_2^*F_4) -
\cos\theta (\mid F_1\mid^2+\mid F_2\mid^2) \}
\nonumber\;,\\
R_T T_{x'} &=& -\sin^2\theta\,Re\{\frac{\cos\theta}{2}( \mid F_3\mid ^2 +
\mid F_4\mid ^2) + F_1^*F_3 + F_2^*F_4 + F_3^*F_4\}
\nonumber\;,\\
R_T T_{z'} &=& \sin\theta\,Re\{\frac{\sin^2\theta}{2}(\mid F_4\mid ^2 -
\mid F_3\mid^2) + F_1^*F_4 - F_2^*F_3 + \cos\theta(F_1^*F_3 - F_2^*F_4)\}
\nonumber\;,\\
R_T L_{x'} &=& \sin\theta\,Re\{\frac{\sin^2\theta}{2}(\mid F_3\mid ^2 -
\mid F_4\mid^2) -  \mid F_1\mid ^2 + \mid F_2\mid^2 +
\nonumber\\ &+&
F_2^*F_3 - F_1^*F_4 - \cos\theta(F_1^*F_3 - F_2^*F_4)\}
\nonumber\;,\\
R_T L_{z'} &=& Re\{2F_1^*F_2 + \sin^2\theta(F_1^*F_3 + F_2^*F_4 + F_3^*F_4) -
\nonumber\\ &-&
\cos\theta (\mid F_1\mid ^2 + \mid F_2\mid^2) +
 \frac{\sin^2\theta}{2}\cos\theta(\mid F_3\mid ^2 + \mid F_4\mid^2) \}\;.
\end{eqnarray*}

\section{expansion in leading multipoles}
Complete expansions of the polarization observables into multipoles
up to $l=1$ can be found in
Refs.\cite{Dre92,Fas92}.\footnote{Note that Ref.\cite{Fas92}
uses a different sign for the observables
$E,H,O_{x'},O_{z'},C_{x'},C_{z'}$ and $L_{x'}$. }
As this is sufficient
for most cases of pion photoproduction, in eta photoproduction also $l=2$
multipoles play an important role even at threshold. However, a general
expansion up to $l=2$ is very extensive and difficult to survey. Therefore,
we give only those leading multipoles that are excited by nucleon
resonances, the $E_{0+}$ by the $S_{11}(1535)$,
the $M_{1-}$ by the $P_{11}(1440)$ and the $E_{2-},M_{2-}$ excited
by the $D_{13}(1520)$. Furthermore, since the $S_{11}$ dominates so
strongly, we restrict ourselves only to those contributions proportional
to $E_{0+}$.

\begin{eqnarray*}
R_T &=& \mid E_{0+}\mid^2 - Re\{E_{0+}^*[ 2\cos\theta M_{1-} -
(3\cos^2\theta - 1)(E_{2-} - 3M_{2-})]\}
\nonumber\;,\\
R_T \Sigma &=& 3\sin^2\theta Re\{E_{0+}^*(E_{2-} + M_{2-})\}
\nonumber\;,\\
R_T T &=& - 3\sin\theta \cos\theta\,Im\{E_{0+}^*(E_{2-} + M_{2-})\}
\nonumber\;,\\
R_T P &=& - \sin\theta Im\{E_{0+}^*[2M_{1-} - 3\cos\theta(E_{2-}- 3M_{2-})]\}
\nonumber\;,\\
R_T G &=& -3\sin^2\theta Im\{E_{0+}^*(E_{2-} + M_{2-})\}
\nonumber\;,\\
R_T H &=& \sin\theta Im\{E_{0+}^*[2M_{1-} -3\cos\theta(E_{2-} - 3 M_{2-})]\}
\nonumber\;,\\
R_T E &=& \mid E_{0+}\mid^2 - Re\{E_{0+}^*[2\cos\theta M_{1-} -
(3\cos^2\theta-1)(E_{2-} - 3 M_{2-})]\}
\nonumber\;,\\
R_T F &=& -3\sin\theta\cos\theta Re\{E_{0+}^*(E_{2-} + M_{2-})\}
\nonumber\;,\\
R_T O_{x'} &=& 3\sin\theta Im\{E_{0+}^*(E_{2-} + M_{2-})\}
\nonumber\;,\\
O_{z'} &=& 0
\nonumber\;,\\
R_T C_{x'} &=&\sin\theta\,[\mid E_{0+}\mid^2 - Re\{E_{0+}^*
(E_{2-} - 3M_{2-})\}]
\nonumber\;,\\
R_T C_{z'} &=& -\cos\theta\mid E_{0+}\mid^2 + 2Re\{E_{0+}^*[M_{1-} -
\cos\theta(E_{2-} - 3 M_{2-})]\}
\nonumber\;,\\
T_{x'} &=& 0
\nonumber\;,\\
R_T T_{z'} &=& -3\sin\theta Re\{E_{0+}^*(E_{2-} + M_{2-})\}
\nonumber\;,\\
R_T L_{x'} &=&-\sin\theta\,[\mid E_{0+}\mid^2 - Re\{E_{0+}^*
(E_{2-} - 3M_{2-})\}]
\nonumber\;,\\
R_T L_{z'} &=& -\cos\theta\mid E_{0+}\mid^2 + 2 Re\{E_{0+}^*[M_{1-} -
\cos\theta(E_{2-} - 3 M_{2-})]\}\;.
\end{eqnarray*}

\begin{figure}
\caption{
Total cross section for the process
$(\gamma,\eta)$ on the proton calculated with PS and PV Born terms.
The full curve contains no Born terms, while the dashed lines are (from
the top down) obtained with $g_{\eta NN}^2/4\pi$=0.1, 0.5, 1.0, and 3.0 for
PS-coupling, and $g_{\eta NN}^2/4\pi$=1.0, 3.0, 6.0 and 10.0 for PV-coupling,
respectively. }
\label{totalpspv}
\end{figure}

\begin{figure}
\caption{
Total cross section for the process
$(\gamma,\eta)$ on the proton.
The full (dashed) curve is obtained with resonances, vector mesons and
PS (PV) Born terms with $g_{\eta NN}^2/4\pi$\
= 0.1 (6.0).  The dotted curve shows the nucleon resonances only while
the dashed-dotted line also includes vector mesons but not Born terms.
The experimental data are from
Ref.\protect\cite{Dyt90} ($\bullet$) and Ref.\protect\cite{AaB68} ($\circ$).}
\label{total}
\end{figure}

\begin{figure}
\caption{Differential cross section for the process
$(\gamma,\eta)$ on the proton.
The curves are as in Fig. 2.
The experimental data are from
Ref.\protect\cite{Dyt90} ($\bullet$) and Ref.\protect\cite{Del67} ($\circ$).}
\label{diff}
\end{figure}

\begin{figure}
\caption{Real and imaginary parts of the $E_{0+}$ multipole for
$(\gamma,\eta)$ on the proton.
The curves are as in Fig. 2. The resonance contribution for this
multipole comes solely from the $S_{11}(1535)$.}
\label{E0+}
\end{figure}

\begin{figure}
\caption{Real part of the $M_{1-}$ multipole for
$(\gamma,\eta)$ on the proton.
The curves are as in Fig. 2. The resonance contribution for this
multipole comes solely from the Roper.}
\label{M1-}
\end{figure}

\begin{figure}
\caption{Recoil polarization for
$(\gamma,\eta)$ on the proton at a photon lab energy of 830 MeV.
The curves are as in Fig. 2. The experimental data point is from
Ref.\protect\cite{Heu70}. }
\label{recoil}
\end{figure}

\begin{figure}
\caption{
Influence of the $P_{11}(1440)$ and $D_{13}(1520)$ resonances on the
differential cross section $d\sigma/d\Omega$, the single-polarization
observables $\Sigma$, $T$ and $P$ and the double-polarization observables
$E,F,G,H$ for polarization of beam and target at a photon lab energy of 752
MeV.
The full lines show the complete calculation with resonances, vector mesons
and PS Born terms with $g_{\eta NN}^2/4\pi=0.4$. The dashed and dotted
lines are obtained when the $D_{13}$ or the $P_{11}$ resonances are
omitted, respectively.}
\label{observables}
\end{figure}

\begin{figure}
\caption{Real part of the $E_{2-}$ multipole for
$(\gamma,\eta)$ on the proton.
The curves are as in Fig. 2.}
\label{M1-}
\end{figure}

\begin{figure}
\caption{Differential cross section for eta photoproduction on p, n, d,
$^3$He, $^3$H and $^4$He. The experimental data on the proton are from
Ref.\protect\cite{Dyt90} ($\bullet$) and Ref.\protect\cite{Del67} ($\circ$),
the data point on the deuteron is from Ref.\protect\cite{And69}
($^{_{\triangle}}$).}
\label{PWIA}
\end{figure}

\begin{table}
\caption{
Parameters for the $\pi N$ and $\eta N$ self-energies in $MeV$}
%\begin{center}
%\renewcommand{\arraystretch}{1.5}    % check for better line separation
\begin{tabular}{cccccccccc}
& & $a$ & $b_1$ & $b_2$ & $c_1$ & $c_2$ & $d_1$ & $d_2$ & $d_3$ \\
\tableline
$S_{11}$ & $\pi N$ & 17 & 0 & 0 & 0 & 0 & -129.5 & 80 & -5 \\
        & $\eta N$ & -27 & 17.7 & -1.23 & 22.9 & -5.17 & -38.1 & 18.3 & 0
\\
$P_{11}$ & $\pi N$ & -150 & 0 & 0 & 0 & 0 & 55.1 & -96.2 & 6.6 \\
$D_{13}$ & $\pi N$ & -26 & 0 & 0 & 0 & 0 & 23 & -32.1 & 2.7 \\
\end{tabular}
%\end{center}
\end{table}

\begin{table}
\caption{
Coupling constants and cut--off masses for
the background vector meson exchange contributions.}
\begin{tabular}{ccccc}
${V}$ & $g_{V}^2/4\pi$ & $g_{T}/g_{V}$ & $\Lambda_{V} (MeV)$ & $\lambda_{V}$
\\
\tableline
$\omega$   &  23  &  0  & 1400 & 0.192 \\
$\rho$     &  0.5 & 6.1 & 1800 & 0.89 \\
\end{tabular}
\end{table}

\begin{table}
\caption{
Contributions to the threshold amplitudes
of $E_{0+}$ in units of $10^{-3}/m_{\pi}$. In the lab frame, the threshold
photon energies are $707.16 MeV$ on protons and $706.94 MeV$ on neutrons
for an eta mass of $547.45 MeV$.
The Born terms have been calculated with $g_{\eta NN}^2/4\pi$=1.0}
\begin{tabular}{cccccc}
target & $S_{11}(1535)$ & $\omega$ & $\rho$ & Born $PS(1.0)$ & Born $PV(1.0)$
\\
\tableline
proton   & 12.91 + 5.97 i & 0.35 &  2.63 & -5.20 & -0.88 \\
neutron  & -7.12 - 4.86 i & 0.35 & -2.63 &  3.55 & -1.04 \\
\end{tabular}
\end{table}

\newpage
\end{document}